# Spreadsheet Risk - A new direction for HMRC?

Don Price
HM Revenue & Customs
Eden House, Chester, CH4 9QY
don.price@hmrc.gsi.gov.uk

ABSTRACT

*Her Majesty's Revenue & Customs (HMRC) was born out of the need to create a UK tax authority by merging both the Inland Revenue and HM Customs & Excise into one department. HMRC encounters spreadsheets in tax-payers' systems on a very regular basis as well as being a heavy user of spreadsheets internally.*

*The approach to spreadsheet risk assessment and spreadsheet audit is by the use of trained computer auditors and data handlers. This, by definition, limits the use of our specialist spreadsheet audit tool to such trained staff. In order to tackle the growing use of spreadsheets, a new way of approaching the problem has been piloted. The aim is to issue all staff who come across spreadsheets with a simple to use analysis and risk assessment tool, based on the departmental software SpACE (Spreadsheet Audit & Compliance Examination).*

1. **BACKGROUND**

Historically, UK tax authorities have been encountering spreadsheets during systems audits at businesses since the inception of spreadsheets; however the risks inherent didn't immediately manifest themselves and it was some time before they became apparent and were ultimately acknowledged. Initially it was computer auditors who, having recognised the existence of these risks, were able to address them using software tools to examine the spreadsheet in detail at file level rather than as hard copies. In fact, Ray Butler's paper, "Is this spreadsheet a tax evader" did much to raise awareness of those risks, and it was subsequently to help prove the problem was common amongst the international community and was not restricted to the United Kingdom.

Spreadsheets used by businesses may serve many purposes, some of which include the monitoring or controlling of both direct or indirect taxes, which in the UK includes taxes such as VAT, Corporation Tax and PAYE. They usually play a major role within the business accounts and will likely be found at the most critical points in the audit trail, significantly impacting upon the revenue base. They are often designed and developed by company employees who are rarely qualified to do so and are seldom checked to make sure they do what was actually intended, not realising that creating a spreadsheet is akin to programming and the untested spreadsheet is as dangerous and untrustworthy as an untested program.

2. **OUR APPROACH**

Having recognised and acknowledged the inherent risks posed by the use of spreadsheets within audit trails, we created our own spreadsheet audit tool, SpACE (Spreadsheet Audit & Compliance Examination), designed, written and developed to our own specification, to allow us to begin examining spreadsheet files in-depth. This provided us with a means to test the accuracy and completeness of those spreadsheets and as expected we found errors.

Obviously, from the thousands of spreadsheets HMRC encounters, realistically only a small proportion can be examined in depth by our specialist auditors using SpACE. To help us decide which spreadsheets to target, we use our own methodology, naturally risk based – i.e., how much potential tax revenue is at risk? This question will then be followed by others about the spreadsheet itself, its design, development, author, documentation and place in the audit trail, all of which will inform the risk process.

Some spreadsheets reviewed belong to stock or accounting datasets which are more suitably examined using file interrogation software, such as IDEA or ACL, rather than carrying out a full spreadsheet audit using SpACE.

As mentioned previously, should a full spreadsheet audit using SpACE be warranted, then this will be carried out by specially trained staff, most of whom are computer auditors, data handlers or internal auditors.

SpACE of course is widely established and well known to colleagues within EuSpRIG. For more information on obtaining a copy for either evaluation or purchase, please see the Lexis Nexis website at http://rimer.butterworths.co.uk/webcat/enquiry/index.htm and search for "SpACE".

## 3.  EFFECTIVENESS

In order to assess how effective we have been, for the past three years I have carried out a survey of SpACE usage, using the months January to June as a baseline for each year. The survey, sent to SpACE equipped staff, captured facts and figures relating to our spreadsheet audit work, to ensure that SpACE is being used consistently and to disclose any figures relating to errors found and revenue affected.

These annual surveys have centred on SpACE usage by all SpACE equipped staff within HMRC during 2003 to 2005. However, due to different working practices between different parts of HMRC, I have limited the published results to the replies from SpACE equipped staff within what is now the Specialist Unit.

The staff surveyed provided information about all spreadsheet audit activity undertaken during January to June inclusive each year to give a standard half years' worth of data. All questionnaires issued were followed up by me to ensure a 100% response rate. As details of spreadsheet audits are not captured separately from normal records within HMRC, I have had to rely on staff checking their own audit records to extract details of SpACE audits.

I know from my own experience of attending EuSpRIG conferences in the past that facts and figures relating to spreadsheet audits and findings are of particular interest.

In order to understand the terms used in the table below:

>1. The sample size is the number of staff surveyed.
>2. The tax revenue throughput is the amount of tax that is calculated or controlled using the spreadsheets and by default is a fraction of the true amount of money going through most of the spreadsheets.

3. All audits are quantified by the amount of time they took to complete in hours, which is recorded by our systems.
4. The tax assured per audit hour is calculated from the tax throughput and the total hours spent on spreadsheet audit.
5. The number of workbooks and worksheets examined is taken from the individual audit files held by staff.
6. The average hours spent should be looked at in the context of a 37 hour working week for HMRC staff and includes testing, quantifying and reporting the results.
7. The "error rate" percentage is based on the number of audits performed against the number of audits which resulted in additional revenue being identified.

NB: During the periods of the survey, no spreadsheets were audited which resulted in revenue being overpaid to the Department, all quantified errors resulted in money being due to the Department.

**Survey Results**

| Item | Jan – June 2003 | Jan – June 2004 | Jan – June 2005 | Average 2003 – 2005 |
| --- | --- | --- | --- | --- |
| Sample size | 203 | 314 | 255 | 257 |
| Total tax revenue throughput on all Spreadsheets audited | £7.47 billion | £2.75 billion | £3.47 billion | £4.56 billion |
| Time spent on audits in hours | 2,118 | 1,202 | 1,320 | 1,546 |
| Tax assured per audit hour | £3.53 million | £2.29 million | £2.63 million | £2.95 million |
| No. workbooks audited | 321 | 257 | 188 | 255 |
| No. worksheets audited | 1669 | 1355 | 973 | 1,332 |
| Average hours spent per audit | 34.7 | 21.5 | 38.8 | 31.7 |
| Average % error rate per audit (audits producing additional revenue) | 13% | 16% | 12% | 14% |

**4. DEVELOPMENT OF SPACELITE**

The surveys conducted did more than just produce figures for analysis. They identified a need that has led directly to a change of direction in spreadsheet risk assessment, with the development of a spreadsheet risk assessment tool to complement SpACE, which is now being piloted within HMRC.

Spreadsheets are now so prevalent in business that billions of pounds of tax revenue flow through them each year. HMRC has a very large staff base, which services a huge customer base of millions of potential spreadsheet-using taxpayers. It is obvious therefore that the more staff that can be equipped to carry out some form of limited risk assessment and analysis of spreadsheets, the more tax assurance can be gained.

However, SpACE is a complex tool and requires a heavy training commitment, with users needing a higher level of knowledge of Excel than most staff, to understand its output. With the best will in the world, the Department cannot afford the training time needed in order to train everybody in the use of SpACE and spreadsheet audit techniques. In addition, no matter how determined we are in identifying, risk assessing and auditing spreadsheets, it is physically and financially impossible for us to audit every spreadsheet, even if every member of staff was equipped and trained to do so, particularly with an average spreadsheet audit taking up to a week to complete.

It is therefore essential that we target our resources, including our use of SpACE, on those spreadsheets which we consider pose the greatest revenue risk. Therefore, to complement SpACE; we have designed a version for intended use by all staff. SpACELite, as it is called, enables its occasional user to risk-assess and perform a limited range of checks and analysis on spreadsheets they routinely encounter. This has been piloted, very successfully and has led to a far wider appreciation of spreadsheet risks by the users, extra revenue and tax assurance plus, as a by-product, a greater understanding of the workings of Excel. The intention is to make SpACELite more widely available to staff across HMRC and is aimed at providing both analysis of traders' spreadsheets and increased use to examine and test our own internal spreadsheets.

**5. WAY FORWARD**

I would like to see SpACELite eventually made available to every member of the department who needs it.  SpACE users will be able to concentrate their attention on the riskier, more complex and sizeable spreadsheets that warrant their attention. SpACELite users will be able to build upon their own knowledge allowing them to both risk assess and perform analyses on more spreadsheets, providing better tax assurance, but will be able to feed riskier spreadsheets through to the SpACE users for more in-depth analysis. This includes both external and internal spreadsheets.

It is important to note that we perceive a difference between external and internal spreadsheets.

External spreadsheets (from tax payers) are audited to the extent that the "bottom line", i.e. the figure which calculates the tax revenue, is correct.  We are not overly interested in weaknesses or errors of formatting, formulae, etc, if they do not affect this figure, although such failings will provide some indication of the overall credibility of the business.  We will always report such errors as findings but, in the main, will only take action when the tax figures are wrong – this is the only type of error recorded on our systems.

Internal spreadsheets are dealt with differently – we control them and we expect them to be accurate, as often they will be used to make business critical decisions. Like most businesses, HMRC makes many major decisions, such as those relating to Human Resource matters, management accounting, budgets, etc., based on spreadsheet outputs. In such cases, all weaknesses are reported with appropriate recommendations regarding layout, formatting, formulae, etc.

It is one of my aspirations of the SpACELite project to include the tool as part of all internal spreadsheet development protocols, so that spreadsheets are rigorously tested prior to their acceptance.  This will not stop them being examined in the future, but should reduce the risks in the early stages of their use.

As a direct result of the publicity surrounding the development and testing of SpACELite, I have been asked to examine more internal spreadsheets in the past six months than in the preceding three years. The enquirers for these audits have also been issued with SpACELite as part of the pilot programme.

## 6. CONCLUSION

HMRC are in a unique position to be able to examine spreadsheets relating to all types of business, both national and international and to test them for accuracy and completeness. We are able to demonstrate that both fundamental weaknesses and material errors are commonly found in most spreadsheets seen by our staff. The majority of the spreadsheets received for taxation purposes are held by professionals in the business world, and are often to be found in the audit trail for signed documents attesting to the accuracy of business tax returns.

By providing all staff with a simple means of risk assessing and analysing spreadsheets, it is hoped that external spreadsheets will be examined in far greater numbers than have previously been possible, thus leading to greater revenue assurance benefits and the accuracy of internal spreadsheets will also improve.

Any views expressed in this paper are those of the author and will not necessarily represent those of HMRC.

Blank page